\documentclass[%
 reprint,
superscriptaddress,
 amsmath,amssymb,
 aps
]{revtex4-2}

\usepackage{graphicx}
\usepackage{dcolumn}
\usepackage{bm}
\DeclareMathAlphabet{\mathpzc}{OT1}{pzc}{m}{it}

\usepackage[caption=false]{subfig}
\usepackage{amssymb}
\usepackage{amsmath}
\usepackage{graphicx,bm}
\usepackage{verbatim}
\usepackage{xcolor, soul}
\sethlcolor{yellow}

\usepackage{float}

\begin{document}


\title{{\color{black}SIS model of} disease extinction on heterogeneous directed population networks}

\author{Elad Korngut}
 \email{elad.korngut@mail.huji.ac.il}
 \affiliation{
 Racah Institue of Phyiscs, Hebrew University of Jerusalem, Jerusalem 91904, Israel}
\author{Jason Hindes}
\affiliation{%
 U.S. Naval Research Laboratory, Code 6792, Plasma Physics Division, Washington, D.C. 20375, USA }%
\author{Michael Assaf}\email{michael.assaf@mail.huji.ac.il}%
\affiliation{%
 Racah Institue of Phyiscs, Hebrew University of Jerusalem, Jerusalem 91904, Israel}
%


\begin{abstract}
Understanding the spread of diseases through complex networks is of great interest where realistic, heterogeneous contact patterns play a crucial role in the spread. Most works have focused on mean-field behavior -- quantifying how contact patterns affect the emergence and stability of (meta)stable endemic states in networks. On the other hand, much less is known about longer time scale dynamics, such as disease extinction, whereby inherent process stochasticity and contact heterogeneity interact to produce large fluctuations that result in the spontaneous clearance of infection. Here we show that heterogeneity in both susceptibility and infectiousness (incoming and outgoing degree, respectively) has a non-trivial effect on extinction in directed contact networks, both speeding-up and slowing-down extinction rates depending on the relative proportion of such edges in a network, and on whether the heterogeneities in the incoming and outgoing degrees are correlated or anticorrelated. In particular, we show that weak anticorrelated heterogeneity can increase the disease stability, whereas strong heterogeneity gives rise to markedly different results for correlated and anticorrelated heterogeneous networks. All analytical results are corroborated through various numerical schemes including  network Monte-Carlo simulations.


\end{abstract}

\maketitle


\section{\label{sec:Intro}Introduction}
Understanding the dynamics of infectious processes is important both for public health and for basic science \cite{background_inf_models,anneald_net}. For the former, epidemic models can  provide control strategies for minimizing disease spread within populations \cite{jason-ira-paths,book_nasell}. For the latter,  general insight can be gained on, e.g., contagion dynamics, which can be applied to other areas from election dynamics to the spread of computer viruses \cite{doi:10.1137/19M1306658,Billings2002261}. A common approach for modeling epidemic dynamics is to partition a population into different compartments and describe the contagious processes using flux terms between various compartments.

A  wide range of compartmental models have been studied in the literature \cite{background_inf_models}, all of which are designed to capture a particular aspect of disease dynamics. In this work, we are interested in endemic dynamics, where infection lingers within a population, past the point of an initial outbreak \cite{pre_miki_2010,miki_2017}. The  simplest model for endemic disease dynamics is described by the SIS model \cite{anneald_net,book_nasell}, where the population is divided into susceptible (S) and infected (I) individuals; an infected individual that interacts with a susceptible has a chance to transmit the disease, making the susceptible infected; conversely the infected individual can recover and become susceptible once again, according to some prescribed probability.

At the deterministic (mean-field) level, the SIS model has two equilibrium points: an endemic (stable) state and an absorbing (unstable) extinct state \cite{jason-ira-paths}. Due to demographic noise emanating from the discreteness of individuals and stochasticity of the reactions, it is possible that a large fluctuation will bring the system from the stable endemic state to the absorbing extinct state \cite{pre_miki_2010,dykman_wkb_exp,ovaskainen_2001,book_nasell}. This noise-driven rare event is only one manifestation of the so-called noise-driven escape, which also includes switching between metastable states (see, e.g., Refs.~\cite{gene_switch,electric_network}), stochastic fade-out \cite{biomolcul} and unusually small or large (extreme) outbreaks \cite{hindes2022outbreak}.

In most cases, such rare events in population dynamics models are considered within a well-mixed or homogeneous setting, where individuals interact with an equal number of neighbors with uniform transition rates. Here, analytical treatment is possible using standard techniques \cite{wkb_miki,ovaskainen_2001}, or, for higher-dimensional systems, by exploiting timescale separation~\cite{dykman2008disease} or various conservation laws~\cite{hindes2022outbreak}. However, it is known that in heterogeneous networks, in which nodes have varying incoming and outgoing degrees or transition rates, the emergence of an endemic state can be dramatically affected \cite{jason-ira-paths,miki-jason.123.068301,pmid29476196}. {\color{black}This is evident in world trade, health care, and social networks, where heterogeneity in the network connectivity was shown to strongly influence disease spread  \cite{Paini2015,King2012-za,Leventhal2015-fb}}.

Rigorously dealing with rare events in heterogeneous networks is highly challenging. Recently, progress has been made in analyzing rare events in networks close to bifurcation, where the dimensionality is reduced \cite{jason-ira-paths}. In other works, disease extinction in the realm of the SIS model was studied on simplified network topologies~\cite{fagnani2017time}, or on small networks using exact calculations~\cite{holme2018epidemic}.
More recently, the mean time to extinction (MTE) was analytically studied on directed networks with partial heterogeneity, either in the incoming or outgoing degree~\cite{pmid29476196,clancy_same_paper}. Here, it was also suggested that, in the context of extinction in the SIS model, topologically \emph{homogeneous} networks with \emph{heterogeneous} transition rates (susceptibility and infectiousness) are equivalent to topologically \emph{heterogeneous} directed networks with \emph{homogeneous} transition rates. Furthermore, in Ref.~\cite{miki-jason.123.068301} the MTE in the SIS model was studied on undirected degree-heterogeneous networks with weak degree dispersion, within an annealed network approximation. Here, the authors have shown that the degree dispersion suffices to determine the MTE in the leading order.
Yet, rigorously dealing with the full extent of heterogeneity, both in the incoming and outgoing degree, and computing the MTE in this case, has not been carried out in the literature so far.

In this work we extend the formalism developed in Refs.~\cite{pmid29476196,miki-jason.123.068301} and use a semi-classical approach to compute the MTE in the SIS model, for fully heterogeneous, directed networks with arbitrary dispersion. We show that correlation between the population network's incoming and outgoing degrees can greatly affect the MTE: correlation in the degrees  brings about a decrease in the MTE, while weak anticorrelation can dramatically increase the MTE. We also compute perturbatively the effect of strong heterogeneity, and show that here too, correlation and anticorrelation in the incoming and outgoing degrees gives rise to markedly different behaviors. 

Our paper is organized as follows. In Sec.~\ref{sec:theory} we introduce a model employing a topologically homogeneous network with heterogeneous transition rates, and establish its equivalence to degree-heterogeneous networks with homogeneous rates, under the annealed network approximation. We then analyze homogeneous networks with bimodal transition rates and explore such networks numerically in Sec.~\ref{sec:num_method}, and analytically in Sec.~\ref{sec:results}, for both weak and strong heterogeneity. Finally, in Sec.~\ref{sec:discussion} we conclude our results and discuss possible generalizations.

\section{\label{sec:theory}Theoretical formulation}
\subsection{Model}
We begin by formulating the SIS model on a topologically homogeneous network with heterogeneous transition rates. Following the notation of Refs.~\cite{pmid29476196,clancy_pearce}, we assume an isolated population of $N$ individuals that is divided into $k$ groups, with group $i$ $(i=1,2,...,k)$ consisting of $N_{i}$ individuals. We denote $f_{i}=N_{i}/N$ the proportion of the population in group $i$, where $\sum_{i}f_{i}=1$. Each group is comprised of susceptible individuals, $S$, and infected individuals, $I$. The tendency of an infected individual from group $i$ to infect its neighbors is referred to as its infectiousness, $\lambda_{i}$, whereas susceptibility,  $\mu_{i}$, measures the tendency of an individual from group $i$ to become infected due to an infected neighbor. Thus, the individual's infection and recovery rates of  group $i$ are
\begin{equation}
\label{eq:transition_rate}
    I_{i}\xrightarrow{W_{+}(I_{i})} I_{i}+1 , \quad
    I_{i}\xrightarrow{W_{-}(I_{i})} I_{i}-1,
\end{equation}
where $W_{+}(I_{i})=(\beta/N)\left(\sum_{j=1}^{k}\lambda_{j}I_{j}\right)\mu_{i}(N_{i}-I_{i})$ is the infection rate, which depends on the   susceptibility of group $i$ and the collective infectiousness, while $W_{-}(I_{i})=\gamma I_{i}$ is the recovery rate.   In addition, $\beta$ and $\gamma$ are overall measures of infectiousness and recovery rates, N is the number of nodes (individuals) in the network and $I_{i}$ is the number of infected individuals in group $i$ \cite{pmid29476196,clancy_pearce}.

Without  loss of generality we can scale $\lambda_{i}$ and $\mu_{i}$ such that the average infectiousness $\left \langle\lambda\right\rangle$ and average susceptibility $\left\langle\mu\right\rangle$ satisfy: $\left \langle\lambda\right\rangle=\sum_{i}\lambda_{i}f_{i}=1$ and $\left\langle\mu\right\rangle=\sum_{i}\mu_{j}f_{j}=1$. Defining the fraction of infected in group $i$ by $y_{i}=I_{i}/N$, using rates~(\ref{eq:transition_rate}), and ignoring demographic fluctuations, the corresponding mean-field rate equations for the average fractions of infected read:

\begin{equation}
\label{eq:rate}
    \dot{y}_{i}=\beta\left(\sum_{j=1}^{k}\lambda_{j}y_{j}\right)\mu_{i}(f_{i}-y_{i})-\gamma y_{i}.
\end{equation}
Equation~\eqref{eq:rate} has an unstable extinction state, $y_{i}=0$, and a stable endemic state, $y_{i}^{*}$, which depends on the transition rates. Notably, the endemic state exists when the basic reproduction number, $R_0$, defined by~\cite{NGM_constuct,NGM_theory,pmid29476196} 
\begin{equation}\label{R0}
R_{0}=(\beta/\gamma)\sum_{i=1}^{k}\lambda_{i}\mu_{i}f_{i},
\end{equation}
is greater than $1$.

Yet, demographic noise and the fact that the extinct state is absorbing renders the stable fixed point metastable \cite{pre_miki_2010,book_nasell}. Accounting for such noise, the master equation for $P(\mathbf{I},t)$: the probability to find at time $t$ infected groups $\mathbf{I}=\{I_{1},...,I_{k}\}$ reads 
\begin{eqnarray}
        &&\frac{\partial P(\mathbf{I},t)}{\partial t}=\sum_{j=1}^{k}\left[W_{-}(I_{j}+1)P(\mathbf{I}+\mathbf{1}_{j},t)-W_{-}(I_{j}) P(\mathbf{I},t) \right. \nonumber\\
        &&\left.+W_+(I_{j}-1)P(\mathbf{I}-\mathbf{1}_{j},t)-W_+(I_{j})P(\mathbf{I},t)\right],
    \label{eq:master_generic}
\end{eqnarray}
where an increase and decrease by one of group $I_{j}$ is denoted by $\mathbf{I}\rightarrow\mathbf{I}\pm\mathbf{1}_{j}$~\cite{dykman_wkb_exp}. Solving this master equation is, in general, analytically impossible, due to the high dimensionality and complex coupling between the degrees of freedom. Below, we show how to treat this master equation for a homogeneous network with bimodal rates (or bimodal degree distributions). Notably, a generalization to arbitrary symmetric networks, and also weakly-asymmetric networks (with small skewness) is possible, using a similar derivation that appears in Ref.~\cite{miki-jason.123.068301}.

The above formalism holds for well-mixed populations, where the contagion process is assumed to occur between groups with different transition rates. We now show that the latter is equivalent to networks with heterogeneous topology of the incoming and outgoing degrees, under the annealed network approximation. {\color{black}Here, the topology is assumed to vary much faster than the transition rates. This results in a contagion process which is established on an average network, where connections are formed according to degree-dependent probability distributions}~\cite{anneald_net}. In such topologically-heterogeneous networks {\color{black}nodes are categorized into groups, where each group} $i$ has a well-defined incoming degree $d_{in}(i)$ and outgoing degree $d_{out}(i)$, resulting in $k<N$ different groups of nodes. Here, we can denote $\nu$ as the rate at which infection transmits along each edge and by $k_{0}^{(in)}$ and $k_{0}^{(out)}$ the average incoming and outgoing degrees such that $k_{0}^{(in)}$=$k_{0}^{(out)} \equiv k_{0}$ (with no excess edges). As a result, the infection rate of a susceptible node $i$ satisfies  
\begin{equation}
\label{eq:rates_to_degree}
    \frac{\nu}{N k_{0}}\left( \sum_{j=1}^{k}d_{out}(j)I_{j} \right)d_{in}(i)(N_{i}-I_{i}),
\end{equation}
which coincides with $W_+(I_i)$ in Eq.~\eqref{eq:transition_rate} upon choosing  $\beta=\nu k_{0},\mu_{i} =d_{in}(i)/k_{0}$ and $\lambda_{i}=d_{out}(i)/k_{0}$ \cite{pmid29476196}. 

\subsection{The case of bimodal networks}
We now use a toy model of heterogeneity in the form of bimodal networks. We first numerically corroborate the  equivalence between heterogeneity in degree and heterogeneity in rates (for bimodal as well as other networks), and then, rigorously study master equation~(\ref{eq:master_generic}) and the MTE for bimodal susceptibility and infectiousness.

\begin{figure}[h]
    \includegraphics[width=1.0\linewidth]{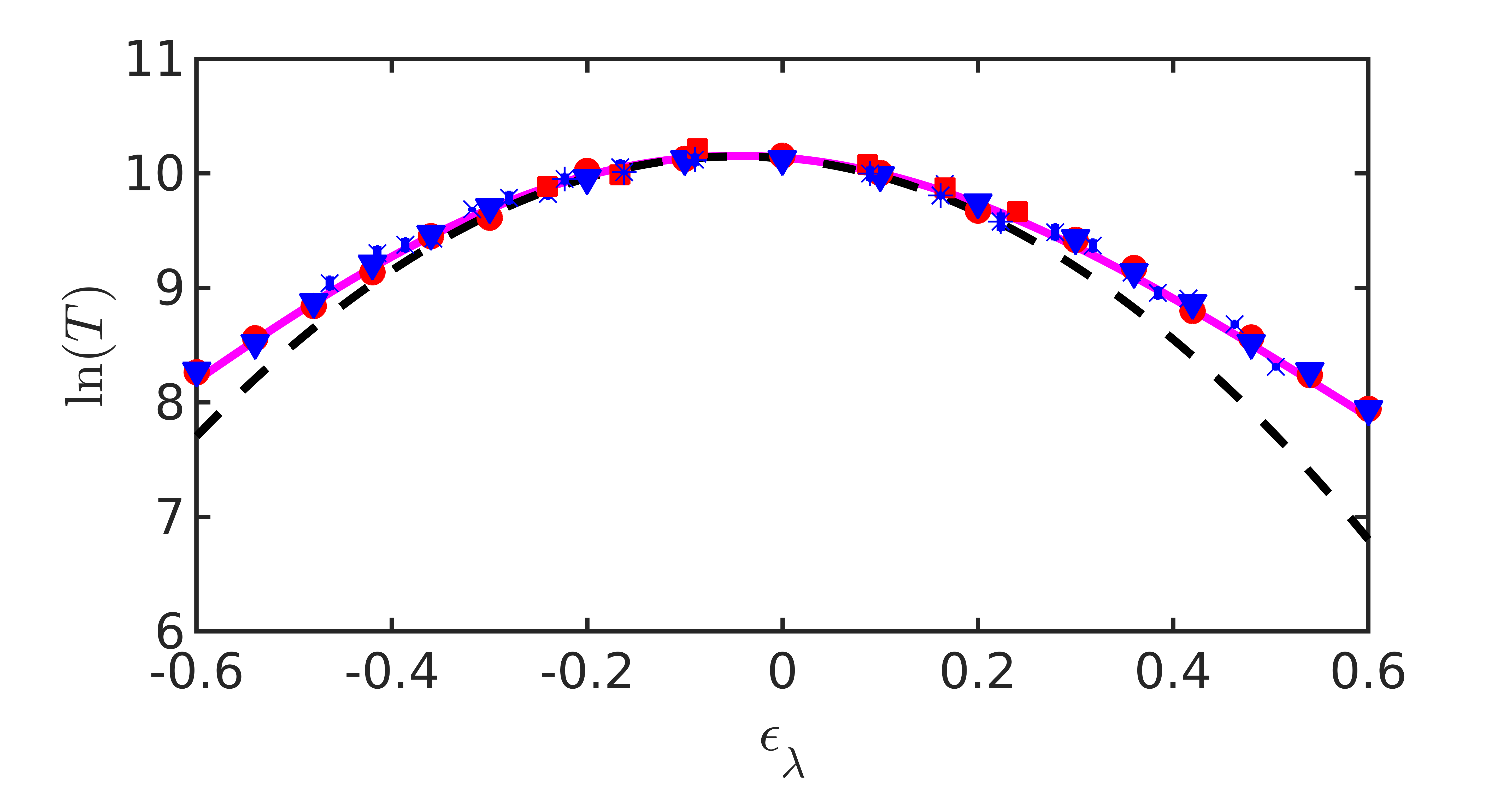}%
\vspace{-3mm}    \caption{The logarithm of the MTE versus $\epsilon_{\lambda}$ for $N=300$, $R_{0}=1.3$, $k_{0}=100$ and $\epsilon_{\mu}=0.1$. Symbols are MC simulations; for each point the MTE and error bar are computed by averaging over 10 networks and 100 realizations according to the scheme presented in Sec.~\ref{sec:num_method}. Results are shown for networks with bimodal (circle) and Gaussian (square) rate distributions, and bimodal (triangle), Gaussian (asterisk) and Gamma (X) degree distributions. {\color{black}Here and henceforth, the size of the error bars is limited by the symbol sizes}. The solid line is the numerical solution of master equation~(\ref{eq:master_generic}), while the dashed line is the solution to Eq.~\eqref{eq:action_both_weak}, see Sec.~\ref{sec:results}.} 
    \label{fig:compare_rate_degree}
\end{figure}

Our starting point is a homogeneous network with heterogeneous rates. Here, for bimodal rates there are two well-mixed interacting populations $I_{1,2}$  with infectiousness $\lambda_{1,2}$, susceptibility $\mu_{1,2}$, and transition rates  given by Eq.~\eqref{eq:transition_rate}, with $k=2$. We now define $\epsilon_{\mu},\epsilon_{\lambda}\in[-1,1]$ as the coefficients of variation (CV) of the infectiousness and susceptibility, respectively, which equal the distributions' standard deviation divided by its mean. In the bimodal case, using the CVs the rates can be rewritten as: $ \lambda_{1} =1-\epsilon_{\lambda}$, $\lambda_{2}=1+\epsilon_{\lambda}$, $\mu_{1} =1-\epsilon_{\mu}$, $\mu_{2} =1+\epsilon_{\mu}
$. Comparing Eqs.~\eqref{eq:transition_rate} and (\ref{eq:rates_to_degree}), one immediately obtains an equivalence with a network of bimodal incoming and outgoing degree distributions with CVs, $\epsilon_{in},\epsilon_{out}\in[-1,1]$, upon choosing $\epsilon_{\mu}=\epsilon_{in}$ and $\epsilon_{\lambda}=\epsilon_{out}$. This can be seen in Fig.~\ref{fig:compare_rate_degree}, where  we  compare the predictions of the master equation for the MTE with Monte-Carlo (MC) simulations, and show the equivalence between degree and rate heterogeneity, for bimodal and other types of networks. The numerical scheme used to obtain Fig.~\ref{fig:compare_rate_degree} is presented in Sec.~\ref{sec:num_method}.  Note that, this figure demonstrates that the MTE is equivalent for  networks with different degree (or rate) distributions, as long as their CV is kept fixed.

Notably, in all our comparisons between different networks and different sources of heterogeneity, we keep the distance from the threshold $R_0$ constant, thereby insuring that the stability of the disease-free (extinct) state remains the same. Using Eq.~(\ref{R0}), keeping a constant $R_0$ requires adjusting the ratio $\beta/\gamma$ for varying susceptibilities and infectiousness; for bimodal networks, this implies maintaining the equality: $\beta/\gamma=R_{0}/(1+\epsilon_{\lambda}\epsilon_{\mu})$.

We now analyze the stochastic dynamics and find the MTE in the bimodal case. Here, rate equations~(\ref{eq:rate}) for the two infected populations, $y_{1}(t)$ and $y_{2}(t)$, satisfy:
\begin{equation}
        \frac{dy_{i}}{dt}=\frac{R_{0}}{1+\epsilon_{\lambda}\epsilon_{\mu}}(\lambda_{1}y_{1}+\lambda_{2}y_{2})\mu_{i}\left(\frac{1}{2}-y_{i}\right)-y_{i},
    \label{eq:rate_eq_bimodal}
\end{equation}
where $i=1,2$. To account for stochasticity, we use master equation~(\ref{eq:master_generic}) with $k\!=\!2$. Notably, even in this simple case an exact solution to the master equation cannot be found in general. Yet, one can use accurate approximation schemes such as the WKB {\color{black}(after Wentzel Kramers and Brillouin)} method~\cite{dykman_wkb_exp} to study large deviations and the MTE. {\color{black}The WKB method utilizes
a small parameter (in our case $1/N\ll 1$) in order to approximately solve the master equation in a singular limit, see below.} Notably, as we will show, the results greatly simplify when heterogeneity is either weak or strong.

To study large deviation in such systems, we assume that after a short time transient the system enters into a long-lived metastable endemic state, and stays there for very long times, on the order of the MTE. {\color{black}This means that, stochasticity causes the metastable probability distribution to slowly 'leak' into the absorbing state. Thus} we write $P(\mathbf{I})=\mathbf{Q}\left(\mathbf{I}\right)e^{-t/T}$ where $T$ is the MTE and $\mathbf{Q}\left(\mathbf{I}\right)$ is the quasi-stationary distribution (QSD) -- the shape of the metastable state. Here the metastable state slowly
decays in time at a rate $1/T$, while simultaneously the extinction probability grows and reaches the value of one at $t\to\infty$~\cite{pre_miki_2010,dykman_wkb_exp}. We now assume that $ N\gg 1 $, plug the metastable ansatz into the master equation [Eq.~(\ref{eq:master_generic})], and employ the WKB approximation for the QSD, $\mathbf{Q}\left(\mathbf{I}\right)\equiv\mathbf{Q}\left(\mathbf{y}\right)\sim e^{-N{\cal S}(\mathbf{y})}$. This results in a stationary Hamiltion-Jacobi equation $H(\mathbf{y},\partial_{\mathbf{y}}{\mathcal{S}})=0$  \cite{dykman_wkb_exp}, where ${\mathcal{S}}(\mathbf{y})$ is the action, and the  Hamiltonian satisfies
\vspace{-5mm}

\begin{equation}
    \label{eq:hamiltionian}
    H\!=\!\frac{\beta}{\gamma}\!\left(\! \sum_{j=1}^{k}\lambda_{j}y_{j}\!\right)\!\!\sum_{i=1}^{k}\mu_{i}(f_{i}\!-\!y_{i})(e^{p_{i}}\!-\!1)+\sum_{i=1}^{k}y_{i}(e^{-p_{i}}\!-\!1),
\end{equation}
with $p_{i}\!=\!\partial_{y_{i}}\mathcal{S}$ being the momentum of group $i$, and $k=2$ for the bimodal case~\footnote{In the simple case of k=1, one recovers the Hamiltonian of the usual well-mixed SIS model, $H=R_0 y(1-y)(e^p-1)+y(e^{-p}-1)$, with $\mathbf{\lambda}=\mathbf{\mu}=1$, and $R_0=\beta/\gamma$ \cite{dykman_wkb_exp}.}. As a result, the Hamilton's equations, $\dot{y_i}=\partial_{p_i}H$ and $\dot{p_i}=-\partial_{y_i}H$, read

\begin{eqnarray}
     \dot{y_{i}}&=&\frac{\beta}{\gamma}\left(\sum_{j=1}^{k}\lambda_{j}y_{j}\right)\mu_{i}(f_{i}-y_{i})e^{p_{i}} - y_{i}e^{-p_{i}},\nonumber\\
     \dot{p_{i}}&=&-\frac{\beta}{\gamma}\lambda_{i}\sum_{j=1}^{k}\mu_{j}(f_{j}-y_{j})\left(e^{p_{j}}-1 \right)\nonumber\\
     &+&\frac{\beta}{\gamma}\left(\sum_{j=1}^{k}\lambda_{j}y_{j}\right)\mu_{i}\left(e^{p_{i}}-1\right)
     -\left( e^{-p_{i}}-1 \right).
\label{eq:hamilton}
\end{eqnarray}

\begin{figure}[h]
    \includegraphics[width=1.0\linewidth]{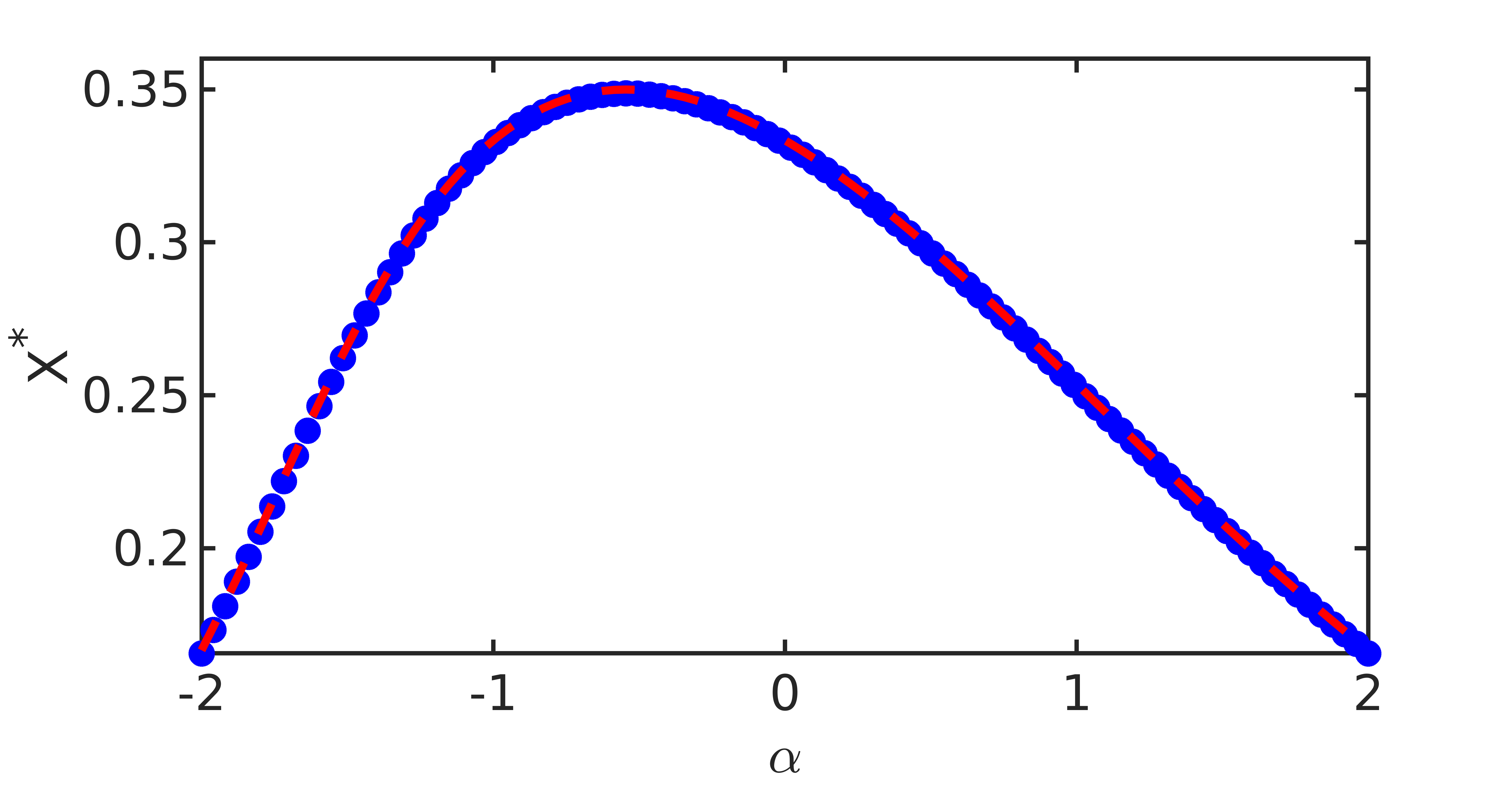}%
\vspace{-4mm}
\caption{The mean population fraction of infected individuals in the endemic state, $X^{*}$, in the bimodal case, as a function of the CVs ratio, $\alpha=\epsilon_{\mu}/\epsilon_{\lambda}$, for $N=2000$, $R_{0}=1.5$ and $\epsilon_{\lambda}=0.5$. Symbols are  numerical solutions to the master equation, while the dashed curve is given by~\eqref{eq:gen_fixed_point}. } 
    \label{fig:master_eq_infected_std}
\end{figure}

The fixed points in the extended phase space $(\{y_i\},\{p_i\})$ can be found, for any degree or rate distribution, by equating  Eqs.~\eqref{eq:hamilton} to zero~\cite{NOLD1980,pmid29476196}. In the bimodal case, $i=1,2$, we find
\begin{equation}
    y_{i}^{*}=\frac{\mu_{i}{D(\epsilon_{\lambda},\epsilon_{\mu})}}{2\left[1+\mu_{i}{D(\epsilon_{\lambda},\epsilon_{\mu})}\right]},\;\;\;\; p_{i}^{*}=-\ln\left[1+\lambda_{i}{D'}\right],
\label{eq:gen_fixed_point}
\end{equation}
where $D(\epsilon_{\lambda},\epsilon_{\mu})=\zeta+\left[\zeta^2+(R_{0}-1)/(1-\epsilon_{\mu}^2)\right]^{1/2}$, $D'$ is identical to $D(\epsilon_{\lambda},\epsilon_{\mu})$ upon replacing $\epsilon_{\lambda} $ and $ \epsilon_{\mu}$, and  $\zeta=R_{0}/[2(1+\epsilon_{\lambda}\epsilon_{\mu})]-1/(1-\epsilon_{\mu}^2)$. Denoting, $X^{*}=y_{1}^{*}+y_{2}^{*}$ as the population fraction of the total number of infected in the endemic state, Eq.~\eqref{eq:gen_fixed_point} with $\epsilon_{\lambda}=\epsilon_{\mu}=0$, agrees with the known homogeneous fixed point, $X^{*}=x_{0}\equiv (R_{0}-1)/R_{0}$. However, for populations where both the susceptibility and infectiousness are heterogeneous, the endemic population can be greater than $x_{0}$ as can be seen in Fig.~\ref{fig:master_eq_infected_std}. Indeed, denoting the CVs ratio by $\alpha=\epsilon_{\mu}/\epsilon_{\lambda}$, for $\alpha>0$ the population's infectiousness and susceptibility are correlated, while for $\alpha<0$ they are anticorrelated. Notably, the maximum of $X^*$ is obtained for anticorrelated rates, which will have a strong impact on the MTE in this regime, see below. 

\subsection{Mean time to extinction}
\vspace{-3mm}
In order to find the MTE, one has to find the action function, $\mathcal{S}(\mathbf{y})$, by solving  Hamilton's equations~(\ref{eq:hamilton}) either numerically or analytically. Below we show how these equations can be solved numerically (Sec.~\ref{sec:num_method}) and analytically (Sec.~\ref{sec:results}). Having found the action function, the MTE in the leading order satisfies~\cite{pre_miki_2010} 
\begin{equation}
\label{eq:def_asym_time}
\begin{aligned}
     T\sim e^{N\Delta\mathcal{S}},\quad\Delta\mathcal{S}=\int_{0}^{\infty}\mathbf{p}\dot{\mathbf{y}}dt=\mathcal{S}(\mathbf{0})-\mathcal{S}(\mathbf{y^{*}}),
\end{aligned}
\end{equation}
where $\Delta\mathcal{S}$ is the action barrier along the optimal path to extinction~\cite{dykman_wkb_exp} and $\mathbf{y}^{*}$ is the endemic fixed point. 

So far, Eq.~\eqref{eq:def_asym_time} has been computed analytically in three special cases. For homogeneous rates, $\epsilon_{\lambda}=\epsilon_{\mu}=0$, one has $\Delta \mathcal{S}_{0}=\ln R_{0}+1/R_{0}-1$~\cite{pre_miki_2010,book_nasell}. In the case of  partial heterogeneity in the susceptibility, one finds~\cite{pmid29476196}
\begin{eqnarray} 
   \hspace{-5mm}\Delta\mathcal{S}(0,\epsilon_{\mu})&=&\frac{1}{2}\ln\left[1+(1-\epsilon_{\mu})D(0,\epsilon_{\mu})\right]\nonumber\\
    &+&\frac{1}{2}\ln\left[1+(1+\epsilon_{\mu})D(0,\epsilon_{\mu})\right]-\frac{D(0,\epsilon_{\mu})}{R_{0}},
    \label{eq:clancy_act_path}
\end{eqnarray}
where $D(\epsilon_{\lambda},\epsilon_{\mu})$ is defined below Eq.~(\ref{eq:gen_fixed_point}), and a similar result holds for partial heterogeneity in the infectiousness.
Finally, in the case of undirected networks, $\epsilon_{\lambda}=\epsilon_{\mu}=\epsilon$, a general result has been found in the case of $\epsilon\ll1$ \cite{miki-jason.123.068301}:
\begin{equation}
\begin{aligned}
\label{eq:mj_action_correction}
    &\Delta\mathcal{S}(\epsilon)=S_{0}-h(R_{0})\epsilon^{2}\\
    &h(R_{0})=\frac{(R_{0}-1)(1-12R_{0}+3R_{0}^2)+8R_{0}^2\ln(R_{0})}{4R_{0}^3}.
\end{aligned}
\end{equation}
Below we generalize these results and solve the Hamilton's equations with heterogeneity present in both rates.

\section{\label{sec:num_method} Numerical simulations}
To corroborate our analytical findings, and to study directed heterogeneous networks in parameter regimes that are inaccessible to analytical treatment, we used three numerical methods to  compute the MTE~\cite{code}. Below, we present these methods  from the most accurate but slowest, to the least accurate but fastest. 

The first method included performing kinetic Monte Carlo (MC) simulations on directed heterogeneous networks. Here, we generated a network satisfying the annealed network approximation and employed Gillespie's algorithm to mimic the SIS dynamics~\cite{config_model,config_molly,doi:gillespie}. For each node in a network there is an exponentially distributed time to make a transition to another state. When the rates are heterogeneous and the topology is homogeneous, each node has a different rate of susceptibility and infectiousness but the same number of neighbors. On the other hand, when the degree distribution is heterogeneous, each node has a different incoming and outgoing degree but the same transition rates, and  the annealed network approximation is used. For correlated networks, degrees (or rates) were paired with the same ranking (the highest  outgoing degree coupled with the highest incoming degree) in contrast with anticorrelated networks, where degrees (or rates) were paired with opposite ranking (highest outgoing degree coupled with lowest incoming degree or vice versa). To account for the variability across different networks, we generated several different network realizations, computed the MTE on each network realization, and then averaged over all networks. For each particular network, the simulated extinction times were fitted to an exponential distribution, where the resulting mean and confidence bounds were treated as the MTE and its confidence bounds of the network. The mean over all networks yielded the overall MTE, whereas the standard deviation of the confidence bounds provided the MTE's error bars. Out of the three methods, this method was the most time consuming {\color{black}with runtime that grows exponentially with increasing reproductive number and population size}; yet it mimics the stochastic dynamics in the most accurate way, and it is applicable to any generic degree or rate distribution.

\begin{figure}[h]
    \includegraphics[width=0.90\linewidth]{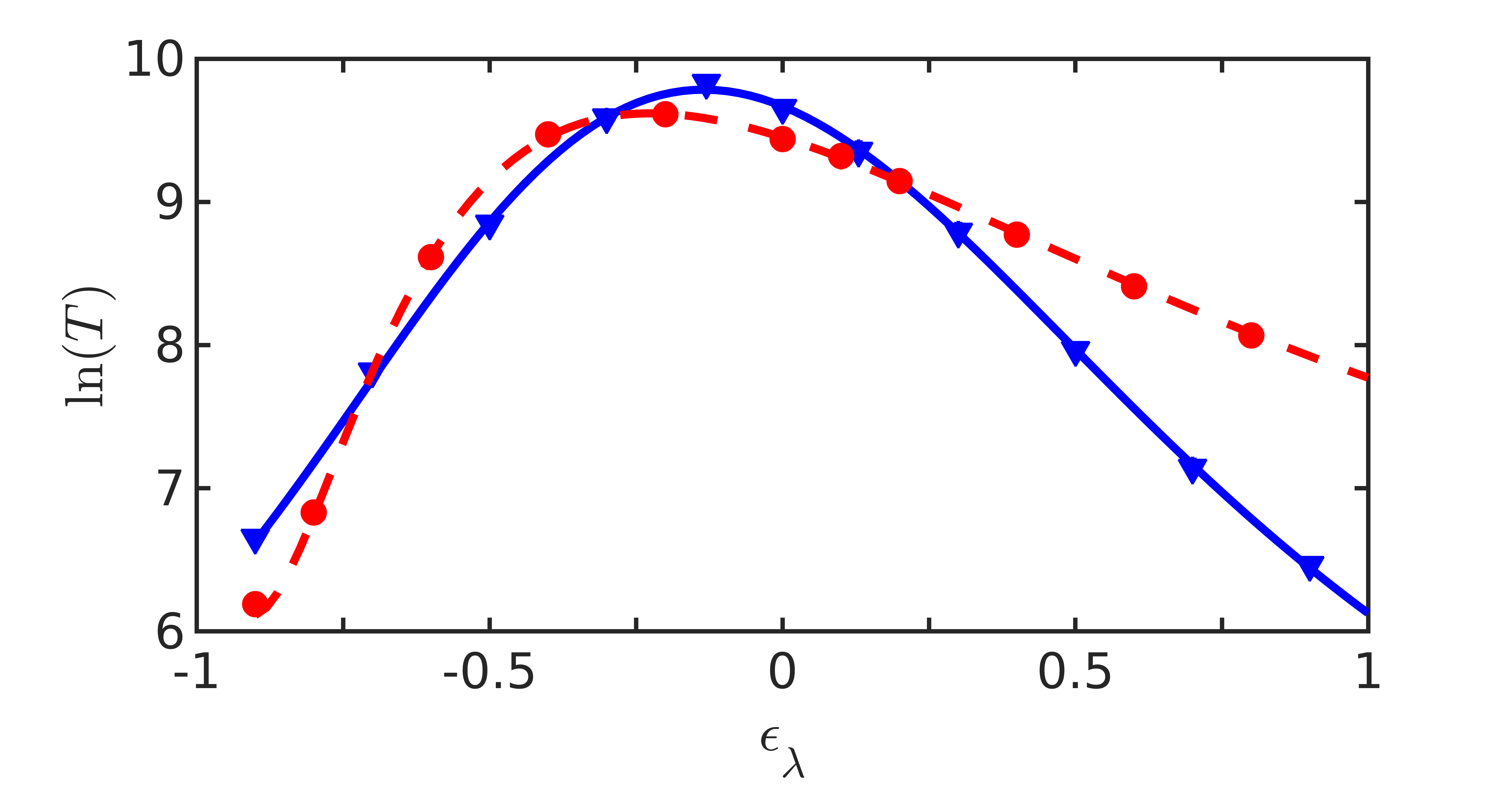}%
\vspace{-3mm}    \caption{The logarithm of the MTE versus $\epsilon_{\lambda}$. 
    Lines are solutions of the master equation, while symbols are MC simulations; for each point the MTE is computed by averaging over 100 stochastic realizations for a fully connected network with a given rate distribution, and then averaging over 10 different network realizations with the same rate distribution. The errors were found to be smaller than the symbol size. Parameters are $N=400$, $R_{0}=1.25$, $\epsilon_{\mu}=0.3$ (triangles) and $N=200$ $R_{0}=1.5$ and $\epsilon_{\mu}=0.8$ (circles).  The maximum MTE was obtained at $\epsilon_{\lambda}=-0.13$ and $\epsilon_{\lambda}=-0.25$, for networks with $\epsilon_{\mu}=0.3$ and $\epsilon_{\mu}=0.8$, respectively.} 
    \label{fig:meqcarlo}
\end{figure}

The second  method included a numerical solution of the master equation, by finding the largest eigenvalue of the matrix in Eq.~(\ref{eq:master_generic}), the inverse of which is the MTE, whereas the corresponding eigenvector represents the QSD~\cite{sim_master}. This method was only used for the bimodal case (k=2){\color{black}; here each population has  anywhere between zero and $N/2$ infected individuals, thus the matrix dimensionality in Eq.~\eqref{eq:master_generic} is $(N/2+1)^2$ for $k=2$.} As shown in Figs.~\ref{fig:compare_rate_degree} and~\ref{fig:meqcarlo}, this method provides highly accurate results for the MTE, including preexponential corrections, unlike the solution to the Hamilton's equations, see below. Furthermore, this method is highly advantageous in terms of running time compared to the MC simulations, as long as $k\leq 2$. For $k>2$, however, this method becomes less feasible when $N$ is large, as one has to deal with a matrix of ${\cal O}(N^k)$ dimensions, and thus, the runtime grows exponentially with increasing $k$.

To demonstrate that the master equation description, which effectively assumes the annealed network approximation~\cite{anneald_net}, can be used to accurately compute the MTE~\cite{sim_master}, we compare  in Fig.~\ref{fig:meqcarlo} its predictions with MC simulations (see also Fig.~\ref{fig:compare_rate_degree}). We do so for bimodal networks, and plot the MTE as a function of $\epsilon_{\lambda}$ for two distinct values of $\epsilon_{\mu}$. Here, the prediction of the master equation for the MTE excellently agrees with that of MC simulations, as long as the average number of neighbors of each node is large~\footnote{In this case, the mean-field like structure of the infection rate in master equation~(\ref{eq:master_generic}) is justified as long as one averages over an ensemble of networks.}.
One can clearly see in Fig.~\ref{fig:meqcarlo} the asymmetry of the curves with respect to the transition point between correlated and anticorrelated degree distributions (i.e., $\epsilon_{\lambda}=0$). Importantly, the maximal MTE is not obtained in the homogeneous scenario, but rather, when the infectiousness and susceptibility are  anticorrelated, see subsection~\ref{subs:both_small}.

The third  method we have used included solving the Hamilton's equations [Eqs.~(\ref{eq:hamilton})] numerically. To do so, we used the iterative action minimization method~\cite{jason-ira-paths,ira_path,lypanov_path}, which allows finding the path $\mathbf{p}(\mathbf{y})$ along the heteroclinic (or zero energy) trajectory, connecting the endemic fixed point $(\mathbf{y},\mathbf{p})=(\mathbf{y^{*}},\mathbf{0})$ with the extinction state $(\mathbf{y},\mathbf{p})=(\mathbf{0},\mathbf{p^{*}})$~\cite{pmid29476196}. Once  $\mathbf{y}(t)$ and $\mathbf{p}(t)$ are found along the optimal path, we use Eq.~\eqref{eq:def_asym_time} to calculate the action. Notably, unlike the MC simulations and numerical solution of the master equation, this method provides the MTE up to exponential accuracy and misses the pre-exponential factor. Yet, it is highly advantageous timewise, as it includes finding solutions to only $2k$ ordinary differential equations {\color{black} for each time-step along the  zero-energy path. Typically, a quasi-Newton method can be used to solve the equations, with quadratic time complexity in the number of time-steps and $k$}. Thus, while we have focused on the bimodal case, this method can be used to deal with any degree (or rate) distribution, with arbitrary $k$ (see details in Ref.~\cite{jason-ira-paths}).

\begin{figure}[h]
\centering
  \includegraphics[width=1.05\linewidth,keepaspectratio]{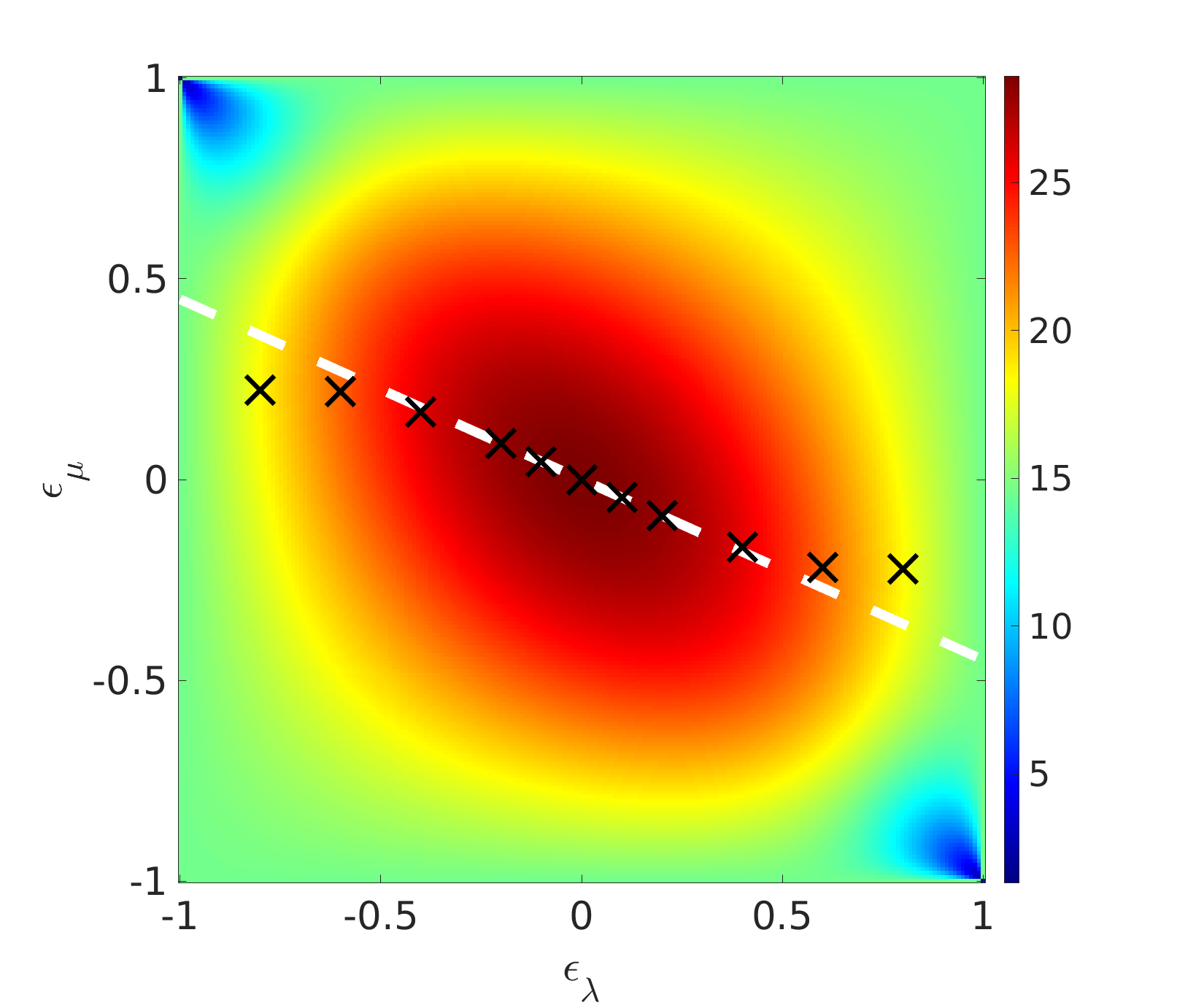}%
\vspace{-5mm}    \caption{A heatmap of the logarithm of the MTE, $\ln(T)$, obtained by solving the master equation in the bimodal case vs $\epsilon_{\lambda}$ and $\epsilon_{\mu}$. Here $N=400$ and $R_{0}=1.5$. The x's mark the value of $\epsilon_{\mu}$ which maximizes the MTE for a given $\epsilon_{\lambda}$, while the dashed line denotes the theoretical prediction for this maximum, see Sec.~\ref{subs:both_small}, formally valid for weak heterogeneity.}
    \label{fig:sim_main}
\end{figure}

Figure~\ref{fig:sim_main} shows numerical solutions of the master equation for the entire phase space in the bimodal case. We notice that, the figure is symmetrical to reflection along either horizontal or vertical axis. This is due to the duality property, which states that for any generic network with SIS dynamics denoted by $\mathbf{T}$, where $T_{ij}$ indicates the rate at which individual j infects individual i, the action barrier remains unchanged under the transposition of matrix $\mathbf{T}$. For our model, $T_{ij}=\beta\lambda_{j}\mu_{i}/N$, which suggests that the action remains the same under the exchange of $\mu$ and $\lambda$ and that, $\Delta\mathcal{S}(\epsilon_{\lambda},\epsilon_{\mu})=\Delta \mathcal{S}(\epsilon_{\mu},\epsilon_{\lambda}) $ \cite{wilkinson_duality,liggett_duality,haris_duality,pmid29476196}.
The rest of the paper will be devoted to finding analytical expressions for $\Delta\mathcal{S}(\epsilon_{\lambda},\epsilon_{\mu})$ in two main regions: weak heterogeneity and strong heterogeneity.

\section{Results\label{sec:results}} 
We now use perturbation theory to analyze two important parameter regions: weak and strong heterogeneity. 

\subsection{Weak heterogeneity \label{subs:both_small}}
Here we analyze the case of weak heterogeneity, where $\epsilon_{\lambda},\epsilon_{\mu}\!\ll\! 1$. In this limit we can Taylor expand $\Delta \mathcal{S}(\epsilon_{\lambda},\epsilon_{\mu})$, the action barrier to extinction, up to second order:
\begin{equation}
\begin{aligned}
    &\Delta \mathcal{S}(\epsilon_{\lambda},\epsilon_{\mu})\simeq\Delta\mathcal{S}(0,0)+\left(\frac{\partial \Delta\mathcal{S}}{\partial \epsilon_{\lambda}}\right)\epsilon_{\lambda}+\left(\frac{\partial \Delta\mathcal{S}}{\partial \epsilon_{\mu}}\right)\epsilon_{\mu}\\
    &+\frac{1}{2}\left( \frac{\partial^{2} \Delta \mathcal{S}}{\partial \epsilon_{\lambda}^{2}}  \right)\epsilon_{\lambda}^{2}+ \left( \frac{\partial^{2} \Delta \mathcal{S}}{\partial \epsilon_{\lambda} \partial \epsilon_{\mu}} \right)\epsilon_{\lambda}\epsilon_{\mu}  + \frac{1}{2} \left( \frac{\partial^{2} \Delta \mathcal{S}}{\partial \epsilon_{\mu}^{2}}  \right)\epsilon_{\mu}^{2},
\end{aligned}
\label{eq:act_o2}
\end{equation}
where all derivatives are evaluated at $\epsilon_{\lambda}=\epsilon_{\mu}=0$. Next, we will show that these terms can be evaluated using Eq.~\eqref{eq:clancy_act_path}, Ref.~\cite{miki-jason.123.068301}, and the duality property.

First, the leading order term satisfies: $\Delta \mathcal{S}(0,0)=\Delta\mathcal{S}_{0}=\ln R_0 + 1/R_0 - 1$, since $\epsilon_{\lambda}=\epsilon_{\mu}=0$ and the problem is reduced to  one dimension~\cite{pre_miki_2010,book_nasell}.

To evaluate the first-order terms with respect to $\epsilon_{\lambda}$ or $\epsilon_{\mu}$, we notice that Eq.~\eqref{eq:clancy_act_path} provides the action when only one of the rates includes heterogeneity. For example, to find $\partial_{\epsilon_{\mu}} \Delta\mathcal{S}$ at $\epsilon_{\lambda}=\epsilon_{\mu}=0$, we can differentiate  Eq.~\eqref{eq:clancy_act_path} with respect to $\epsilon_{\mu}$ and substitute $\epsilon_{\mu}=0$, which yields $\partial_{\epsilon_{\mu}} \Delta\mathcal{S}(0,\epsilon_{\mu})=0$. The same result is obtained for $\partial_{\epsilon_{\lambda}} \Delta\mathcal{S}$ at $\epsilon_{\lambda}=\epsilon_{\mu}=0$, as $\mathcal{S}(0,\epsilon)=\mathcal{S}(\epsilon,0)$ (duality principle). Therefore, both first-order terms vanish in Eq.~(\ref{eq:act_o2}).

We now proceed to computing the second order terms in Eq.~(\ref{eq:act_o2}), where there are three such terms. The derivatives  $\partial_{\epsilon_{\mu}}^{2}\Delta\mathcal{S}$ and $\partial_{\epsilon_{\lambda}}^{2}\Delta\mathcal{S}$ can be computed at  $\epsilon_{\lambda}=\epsilon_{\mu}=0$ in the same way as the first-order terms, using Eq.~\eqref{eq:clancy_act_path}. This yield
$\partial_{\epsilon_{\lambda}}^{2}\Delta\mathcal{S}=\partial_{\epsilon_{\mu}}^{2}\Delta\mathcal{S}=-x_{0}^{2}$~\footnote{The fact that these derivatives are identical stems from the duality principle, namely $\Delta{\cal S}$ remains unchanged under the exchange of $\epsilon_{\mu}$ and $\epsilon_{\lambda}$.}.
To compute the mixed derivative in Eq.~(\ref{eq:act_o2}), we define $\psi(R_0)$ such that $\psi(R_0)x_0^2/2=-\partial_{\epsilon_{\lambda}}\partial_{\epsilon_{\mu}}\Delta\mathcal{S}$, evaluated at $\epsilon_{\lambda}=\epsilon_{\mu}=0$. As a result, the action barrier becomes 
\begin{equation}
    \Delta\mathcal{S}(\epsilon_{\lambda},\epsilon_{\mu})\simeq\Delta \mathcal{S}_{0}-\frac{x_{0}^{2}}{2}\epsilon_{\lambda}^{2}\left[1+\psi(R_0)\alpha+\alpha^{2}\right],
\label{eq:action_both_weak}
\end{equation}
where to remind the reader, $\alpha=\epsilon_{\mu}/\epsilon_{\lambda}$. To find $\psi(R_0)$, we compare Eq.~(\ref{eq:action_both_weak}) with Eq.~(\ref{eq:mj_action_correction}) in the case of undirected networks. Putting $\alpha=1$, we find
$\psi(R_0)=2\left[h(R_{0})-x_{0}^{2}\right]/x_{0}^{2}$, where $h(R_0)$ is given by Eq.~(\ref{eq:mj_action_correction}).

\begin{figure}[h]
    \includegraphics[width=1.07\linewidth]{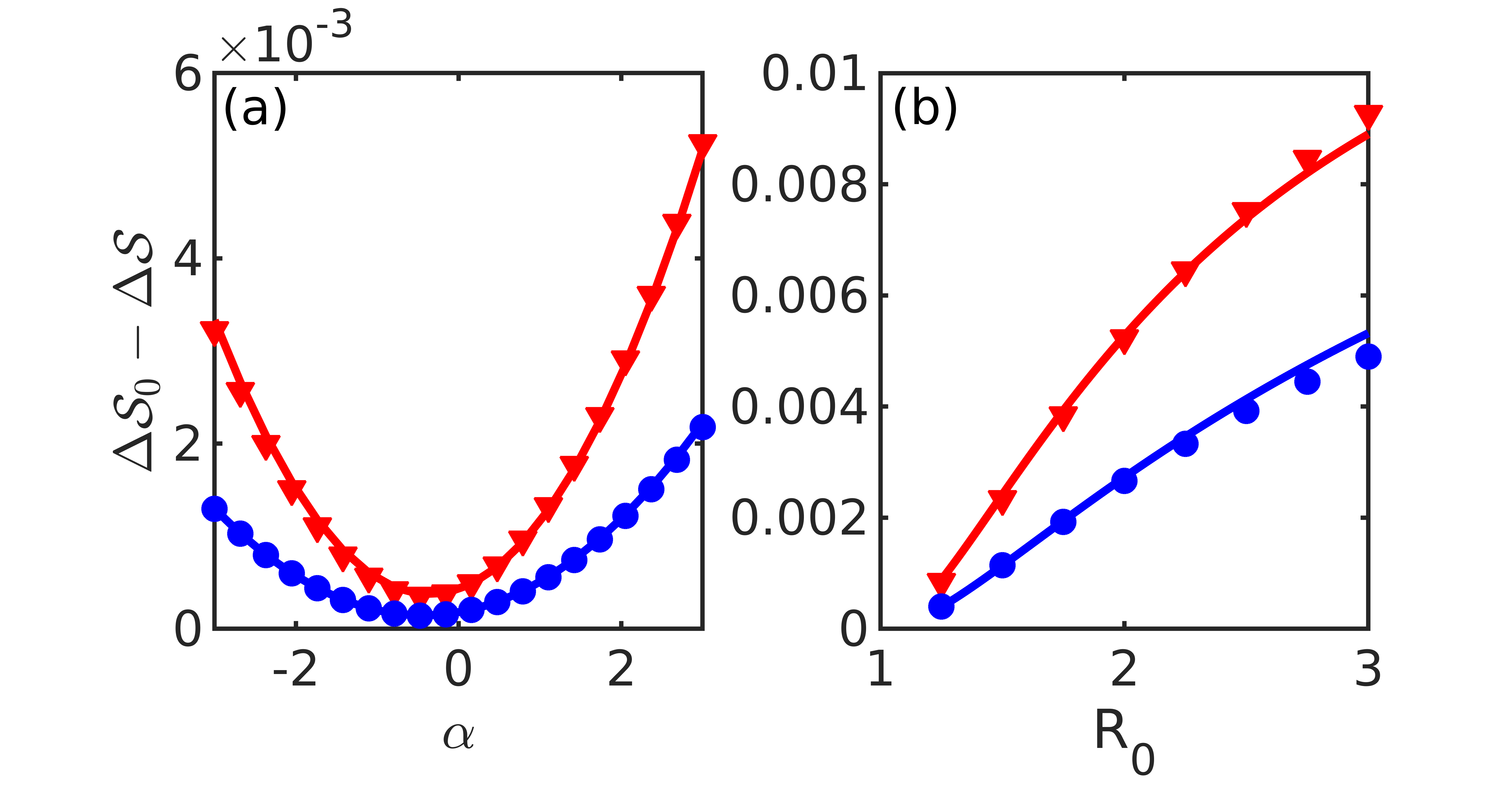}\vspace{-5mm}
    \caption{The correction to the action for weak heterogeneity. Panel (a): $\Delta\mathcal{S}_{0}-\Delta\mathcal{S}$ versus $\alpha$. Symbols are solutions to the Hamilton's equations for $\epsilon_{\lambda}=0.05$, with $R_{0}=1.6$ (circles) and $R_{0}=2.4$ (triangles). Solid curves are the theoretical predictions [Eq.~\eqref{eq:action_both_weak}]. Panel (b): $\Delta\mathcal{S}_{0}-\Delta\mathcal{S}$ versus $R_{0}$. Symbols are solutions to the Hamilton's equations for $\epsilon_{\lambda}=0.08$, with $\alpha=0.5$ (triangles) and $\alpha=-0.5$ (circles). Solid curves are the theoretical predictions [Eq.~\eqref{eq:action_both_weak}].
}    \label{fig:path_cor_h}
\end{figure}

Equation~\eqref{eq:action_both_weak} is our first main result. It generalizes the results of Refs.~\cite{miki-jason.123.068301,pmid29476196} to directed heterogeneous networks, and predicts up to subleading-order corrections the MTE, which only depends on the CV of both rates, and the reproductive number. A comparison between our analytical solution for the MTE, using Eqs.~(\ref{eq:def_asym_time}) and (\ref{eq:action_both_weak}) and a numerical solution of the Hamilton's equations, can be seen in Fig.~\ref{fig:path_cor_h}, and excellent agreement is observed. Importantly, although our derivation has been carried out for bimodal networks, Eq.~(\ref{eq:action_both_weak}) holds for generic networks of arbitrary degree (or rate) distributions, with CVs,  $\epsilon_{\mu}$ and $\epsilon_{\lambda}$, of the incoming (susceptibility) and outgoing (infectiousness) degrees, respectively. This is demonstrated in Fig.~\ref{fig:compare_rate_degree} for various networks, with bimodal, Gaussian and Gamma distributions.

How does $\psi(R_0)$ behave with $R_0$? Close to the bifurcation, at $R_0-1\ll 1$, it can be shown that in the leading order, $h(R_0)\simeq 3/2(R_0-1)^2\simeq 3/2 x_0^2$. As a result, at $R_0-1\ll 1$, $\psi(R_0)\simeq 1$. As $R_0$ increase, the value of $\psi(R_0)$ decreases monotonically. Thus, since $\psi(R_0)\leq 1$ for any $R_0$, the action barrier in the presence of heterogeneous rates is always smaller than that in the fully homogeneous case.  However, given heterogeneity in, say, only the incoming degrees, with homogeneous outgoing degree, the MTE can be \textit{increased} by adding heterogeneity in the outgoing degree as well. This result is counter-intuitive, as in most cases adding heterogeneity increases overall fluctuations, which decrease the stability of the metastable state, and thus, decrease the MTE.  

\begin{figure}[h]
    \includegraphics[width=1.0\linewidth]{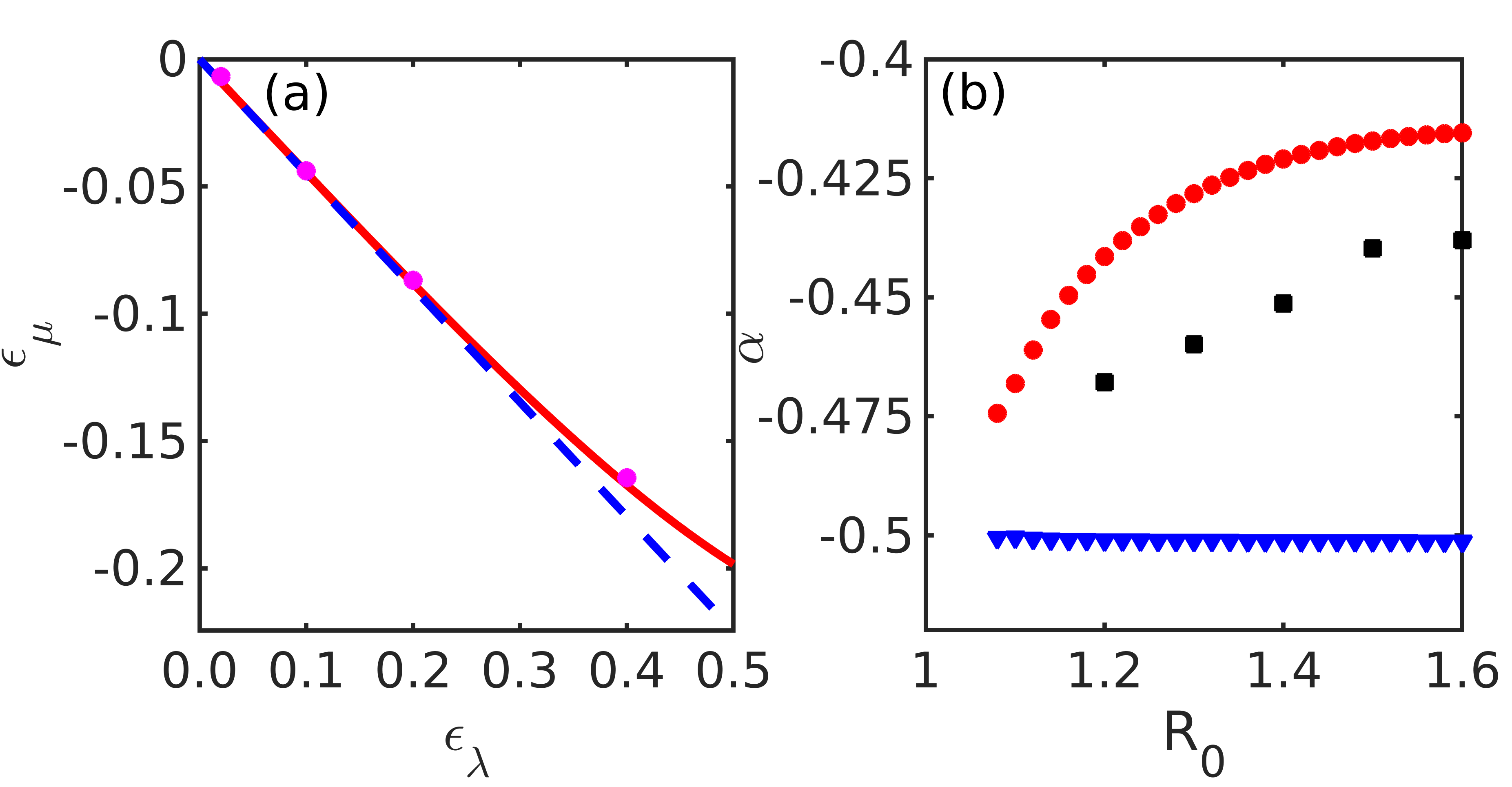}%
\vspace{-4mm}    
\caption{Panel (a): The value of $\epsilon_{\mu}$ for which the MTE is maximized versus  $\epsilon_{\lambda}$. The solid line shows the numerical solution of the master equation, the circles are numerical solutions of the Hamilton's equations, while the dashed line shows Eq.~\eqref{eq:action_both_weak}, valid at  $\epsilon_{\lambda}\ll 1$. Parameters are $N=300$  and $R_{0}=1.5$. Panel (b): the values of $\alpha$ at which: (i) $X^{*}$ receives its maximum (triangles), (ii) the QSD's relative width receives its minimum (circles), and (iii) $\Delta {\cal S}$ receives its maximum (squares), versus $R_0$. The mean and relative width were obtained by numerically solving the master equation with $N=2000$, while the action was obtained by solving the Hamilton's equations. } 
    \label{fig:weak_hetro_meq_lnt_v_eps}
\end{figure}

Indeed, having found the dependence of the action barrier on $\alpha$ for weakly heterogeneous networks, we can find the value of $\alpha$ for which the MTE is maximized. Doing so, we find $\alpha_{max}=-\psi(R_0)/2$, or $\epsilon_{\mu}=-[\psi(R_0)/2]\epsilon_{\lambda}$~\footnote{Here, due to the duality principle, we can exchange $\epsilon_{\lambda}$ and $\epsilon_{\mu}$ and leave the action barrier unchanged.}. For $R_0$ close to $1$, the maximum is obtained at $\alpha=-1/2$, and as $R_0$ increases, the maximum decreases in its absolute value. This behavior can be seen in Figs.~\ref{fig:sim_main} and \ref{fig:weak_hetro_meq_lnt_v_eps}. Notably, the maximum of the MTE is obtained at a negative value of $\alpha$, namely, when the heterogeneity in the incoming degree is anticorrelated with that of the outgoing degree. This is evident also by looking at Fig.~\ref{fig:weak_hetro_meq_lnt_v_eps}b. On the one hand, the mean fraction of total infected is maximized at $\alpha=-1/2$ regardless of the value of $R_0$, which is the dominant factor for the fact that the MTE is maximized at this value of $\alpha$ or close to it.  On the other hand, the minimum of the relative width of the QSD (or relative fluctuations), obtained at  $\alpha\simeq -1/2$ for $R_0-1\ll 1$, shifts towards $0$ as $R_0$ increases. These two effects cause the maximum of the MTE to also shift from $\alpha=-1/2$ towards $\alpha=0$ as $R_0$ is increased. This also indicates that anticorrelation between the incoming and outgoing degrees decreases (for any $R_0$) the typical fluctuations, which brings about an increase in the disease stability and MTE, see Figs.~\ref{fig:sim_main}, \ref{fig:path_cor_h} and~\ref{fig:weak_hetro_meq_lnt_v_eps}.

\vspace{-4mm}
\subsection{Strong heterogeneity}
Here we analyze the case of strong heterogeneity in  either the infectiousness or susceptibility. We denote by $\delta=1-\epsilon$ the distance from maximal heterogeneity, and assume $\delta$ is small. Without loss of generality we study the case of strong heterogeneity in infectiousness, $\delta_{\lambda}=1-\epsilon_{\lambda}\ll1$. Our aim is to compute $\Delta{\cal S}$, given by 
\begin{equation}
\Delta{\cal S}=\int_{y_1^*}^{0}p_1(y_1)dy_1+\int_{y_2^*}^{0}p_2(y_2)dy_2.\label{action_strong}
\end{equation}
In the correlated case, $\alpha>0$, it can be shown that the leading  ${\cal O}(\delta_{\lambda}^0)$ order of Hamiltonian~\eqref{eq:hamiltionian} satisfies
\begin{equation}\label{strcorrham}
    H\!=\!y_{2}\left[ e^{-p_2}-1+(e^{p_2}-1)(2y_{2}-1)R_{0} \right]+\mathcal{O}(\delta_{\lambda}).
\end{equation}
Here, we have assumed that $p_1\sim {\cal O}(\delta_{\lambda})$, see below. Putting $H=0$ we find that $p_{2}^{(0)}(y_{2})=-\ln\left[ R_{0}(1-2y_{2}) \right]$. Integrating over the momentum $p_{2}(y_{2})$ along the extinction path, see Eq.~(\ref{action_strong}), yields $\Delta\mathcal{S}=\Delta {\cal S}_{0}/2$. This is the action obtained for a well-mixed population of size $N/2$, since half of the population has zero infectiousness, thus only half of the population participates in the dynamics.

To obtain the $\delta_{\lambda}$-dependent correction to this result, we compute the fixed points of the Hamilton's equations [Eq.~\eqref{eq:hamilton}], given by Eq.~(\ref{eq:gen_fixed_point}), for $\delta_{\lambda}\ll 1$, which read:
\begin{equation}
\begin{aligned}
    &p_{1}^{*}=-\frac{1}{2}(R_{0}-1)\delta_{\lambda}\\
    &p_{2}^{*}=-\ln(R_{0})-\frac{(R_{0}-1)(1-\epsilon_{\mu})}{2(1+\epsilon_{\mu})}\delta_{\lambda}\\
    &y_{1}^{*}=\frac{x_{0}R_{0}(1-\epsilon_{\mu})}{2\xi}\left[1-\frac{R_{0}(1-\epsilon_{\mu})\epsilon_{\mu}}{\xi^2}\delta_{\lambda}\right]\\
    &y_{2}^{*}=\frac{x_{0}}{2}\left[1+\frac{(1-\epsilon_{\mu})\epsilon_{\mu}}{(1+\epsilon_{\mu})\xi}\delta_{\lambda}\right],
\end{aligned}
\label{eq:fixed_delta_lam}
\end{equation}
where $\xi=R_{0}-(R_{0}-2)\epsilon_{\mu}$. We now proceed in the same spirit as Ref.~{\cite{miki-jason.123.068301}, and assume the following scaling for the momenta, valid for $\alpha>0$: $p_1(y_1)=p_1^*(y_1^*-y_1)/y_1^*$ [which scales as ${\cal O}(\delta_{\lambda})$], and $p_2(y_2)=p_2^{(0)}[y_2(1+\theta\delta_{\lambda})]+[p_2^*+\ln(R_0)](y_2^*-y_2)/y_2^*$, where $p_2^{(0)}$ is given below~(\ref{strcorrham}). In this way, it is guaranteed that $p_1(0)=p_1^*$ and $p_1(y_1^*)=0$. In addition, $p_2(0)=p_2^*$, as $p_2^{(0)}(0)=-\ln(R_0)$, and the free parameter $\theta$ is chosen such that $p_2(y_2^*)={\cal O}(\delta_{\lambda}^2)$, which yields $\theta=(1-\epsilon_{\mu})\epsilon_{\mu}/[\xi(1+\epsilon_{\mu}]$. Using these momenta, and keeping terms up to ${\cal O}(\delta_{\lambda})$, Eq.~(\ref{action_strong}) yields
\begin{eqnarray}
    \label{eq:action_delta_lambda}
    \hspace{-5mm}\Delta\mathcal{S}&=&\frac{\Delta{\cal S}_{0}}{2}+\delta_{\lambda}\frac{1-\epsilon_{\mu}}{4R_{0}(1+\epsilon_{\mu})\xi}\nonumber\\
    &\times&\left[(R_{0}\!-\!1)^{2}R_{0}+(3\!-\!4R_{0}\!+\!R_{0}^2+2R_{0}\ln R_{0})\epsilon_{\mu}\right]\!,
\end{eqnarray}
where the $\delta_{\lambda}$-dependent correction is positive for  $\epsilon_{\mu}>0$. Note that, the case where susceptibility is large and correlated, resulting in small $\delta_{\mu}=1-\epsilon_{\mu}\ll 1$, can be treated using the duality property. In Fig.~\ref{fig:hamilton_eq_high_cor} we show that the analytical formula~(\ref{eq:action_delta_lambda}) excellently agrees with a numerical solution of the master equation for various values of $R_0$ and $\delta_{\lambda}$. Similarly as for weak heterogeneity, Eq.~(\ref{eq:action_delta_lambda}) also holds for generic networks with CVs, $\epsilon_{\mu}$ and $\epsilon_{\lambda}$, of the incoming and outgoing distributions, respectively.

\begin{figure}[h]
    \includegraphics[width=1.0\linewidth]{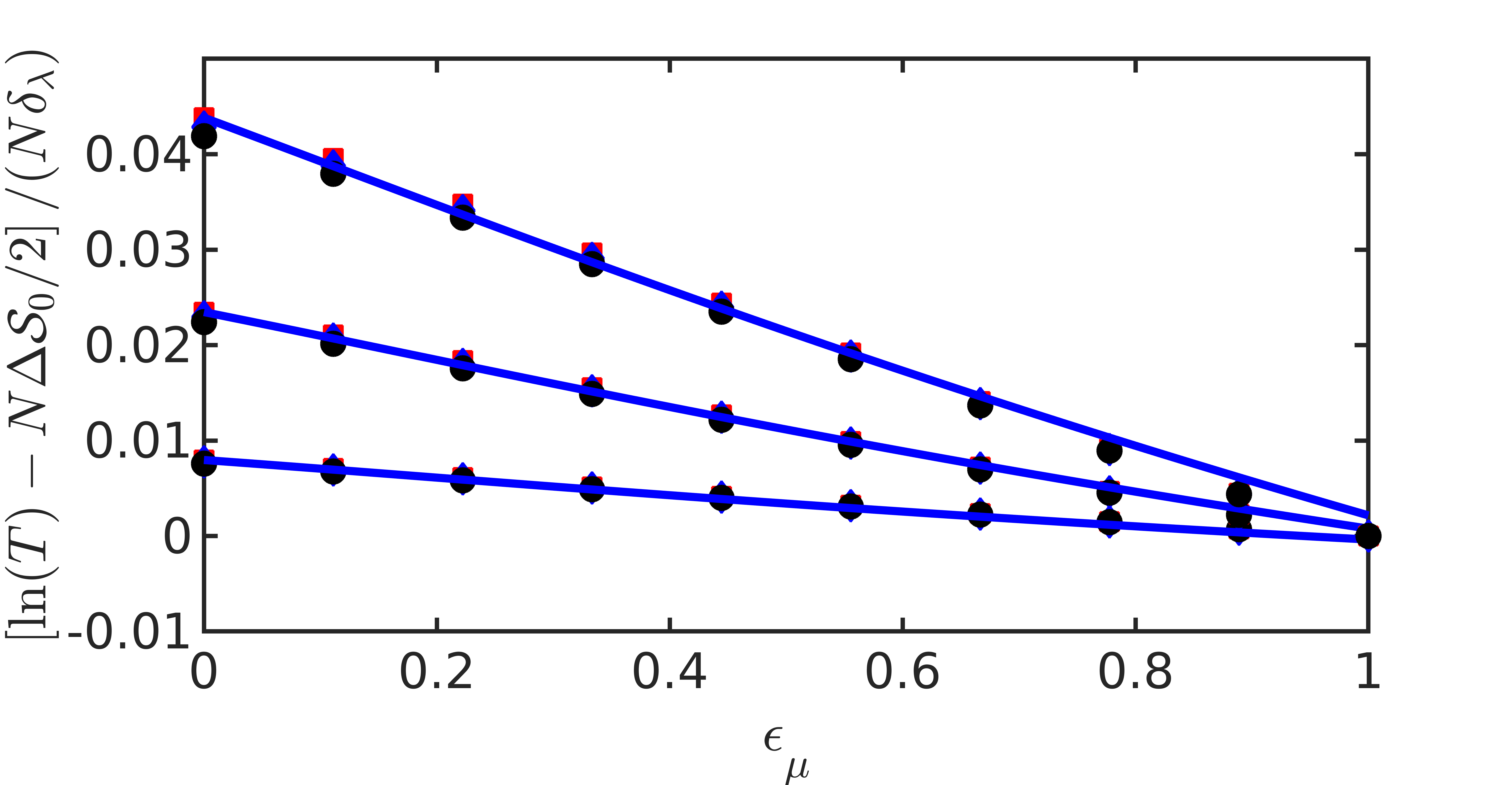}%
\vspace{-4mm}    
\caption{Shown is the correction to the action, $[\ln(T)-N\Delta{\cal S}_{0}/2]/(N\delta_{\lambda})$ as a function of $\epsilon_{\mu}$ for $N=700$. Symbols are numerical solutions to the master equation for $R_{0}=1.2, 1.35, 1.5$ (bottom to top), and for each $R_0$, $\delta_{\lambda}=0.05, 0.1, 0.15$. Lines are the theoretical prediction~(\ref{eq:action_delta_lambda}).} 
    \label{fig:hamilton_eq_high_cor}
\end{figure}

The case of strong anticorrelation, where one of the $\epsilon$'s is close to $1$ and the other is arbitrary and negative, such that $\alpha<0$, is more intricate. Here, the above method is invalid and the result can only be found numerically. 

An interesting example is the case of extreme anticorrelation, with $\epsilon_{\lambda}=1-\delta=-\epsilon_{\mu}$ and $\delta\ll 1$. In this case, in the leading order of $\delta \ll 1$, determinstic rate equations~\eqref{eq:rate_eq_bimodal} yields an unstable extinct state, $y_{1}^{*}=y_{2}^{*}=0$, and a stable endemic state $y_{1}^{*}=(R_{0}-2)/(2R_{0})$ and $y_{2}^{*}=1/2$. This endemic state ceases to exist when $R_0\leq 2$, where only the extinct state exists and becomes stable. That is, bifurcation in this extreme anticorrelation case occurs at $R_0=2$. This is because the first infected group $y_1$ has almost zero infectiousness but high susceptibility, while the second infected group $y_2$ has almost zero susceptibility but high infectiousness. Thus, dynamically, the second group is almost autonomous in the leading order, and becomes established (i.e., reaches an endemic fixed point $y_2^*$) only when the rescaled reproduction number $R_0$ exceeds $2$. Once this occurs, the first infected group can also be established at $y_1^*=1/2$. Only for $R_0\to\infty$ the groups become equivalent. 

\begin{figure}[h]
\includegraphics[width=1.0\linewidth]{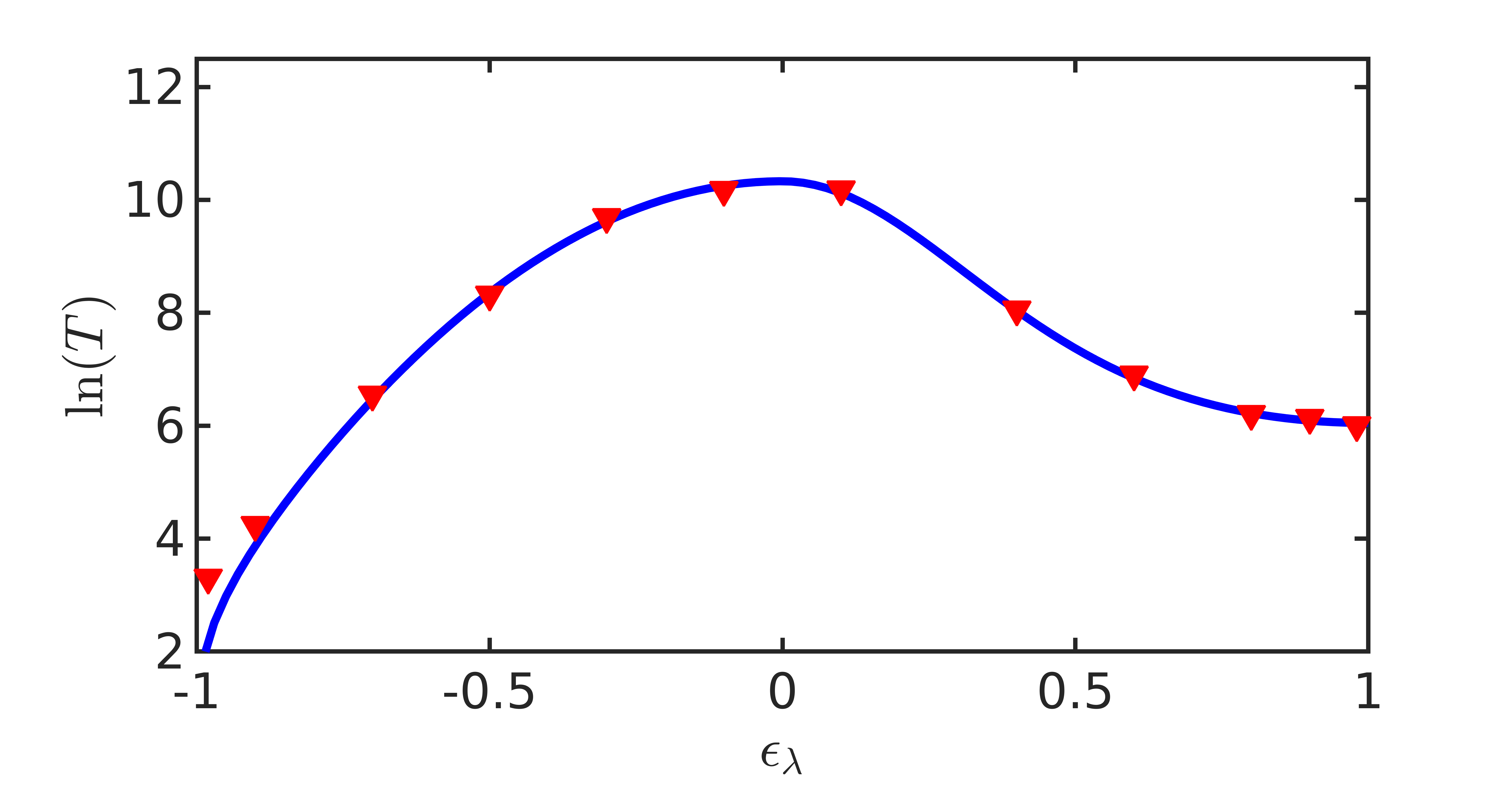}
\vspace{-7mm}
    \caption{The logarithm of the MTE, $\ln(T)$, versus $\epsilon_{\lambda}$ where $\epsilon_{\mu}=\left|\epsilon_{\lambda} \right|$. The solid line is the solution to the master equation for $N=300$ and $R_{0}=1.3$, while symbols are MC simulations obtained for a fully connected network with bimodal rates  averaged over 10 networks and 100 iterations, see Sec.~\ref{sec:num_method}.} 
    \label{fig:eq_miki}
\end{figure}

The MTE in the case of extreme anticorrelation can be numerically computed. In Fig.~\ref{fig:sim_main} this regime can be seen in the top left and bottom right corners when the MTE is minimal, as for $R_0<2$, the endemic state ceases to exist as $\epsilon_{\lambda}=-\epsilon_{\mu}\to 1$ (or vice versa). This can also be observed in  Fig.~\ref{fig:eq_miki}, where we choose the same CV magnitude for both rates, but with a different sign. Here, the extreme antircorrelation case can be seen on the left side where the MTE goes to zero (as the endemic fixed point ceases to exist), whereas extreme correlation can be seen on the right side, where the action approximately converges to half of its maximal value, see Eq.~(\ref{eq:action_delta_lambda}).

\section{Discussion \label{sec:discussion}}
Most studies of disease spread through complex networks focused on short time scales related to  the emergence of the  endemic state and its relaxation dynamics. On the other hand, the study of longer time scales, relevant for disease extinction, was impeded by the complex topology of these networks. Here we have generalized several recent studies, and investigated, in the realm of the SIS model, disease extinction on heterogeneous and directed population networks. To do so we have used various numerical methods with varying efficiency and accuracy, as well as a semi-classical WKB approximation to the master equation. The latter provides, in the leading order in the population size, an alternative Hamiltonian formulation of the problem, which can be dealt with rigorously in various parameter regimes.

We initially showed that under the annealed network approximation, heterogeneity in the network topology, i.e, in the individual's incoming and outgoing degrees, is equivalent to heterogeneity in its susceptibility and infectiousness. In particular, the mean time to extinction (MTE) was shown to be identical for degree and rate heterogeneity, by employing Monte-Carlo  simulations, as well as numerically solving the master equation and the corresponding Hamilton's equations. The numerical schemes were then used to corroborate our analytical findings, mainly in two regimes: weak and strong heterogeneity. While our analytical derivation was carried out on a toy model of bimodal networks {\color{black}with dichotomous heterogeneity, it can be generalized to networks with arbitrary (weakly skewed) incoming and outgoing degree distributions}~\cite{miki-jason.123.068301}. Importantly, we have shown that correlation or anticorrelation between an individual's incoming (susceptibility) and outgoing (infectiousness) degrees, has a dramatic impact on the disease lifetime. It was shown that for strong heterogeneity, anticorrelation tends to always decrease the MTE. On the other hand, for weak heterogeneity, anticorrelation between the incoming and outgoing degrees can increase disease stability and the MTE, which is counter-intuitive, as in most cases heterogeneity tends to decrease stability. 

Although it is beyond the scope of this paper, it would be interesting to explore how asymmetry in the degree (or rate) distribution, and in particular, strongly-skewed distributions such as those with power-law tails, affect the mean time to disease clearance and its statistics.

\section{Acknowledgments}
\vspace{-3mm}
EK and MA acknowledge support from the ISF
grant 531/20. MA also acknowledges Alexander von Humboldt Foundation for an experienced researcher fellowship.

\nocite{*}

\bibliography{hetbib}

\end{document}